# Energy deposition studies for the High-Luminosity Large Hadron Collider inner triplet magnets

N. V. Mokhov,[*] I.L. Rakhno, I.S. Tropin

*Fermi National Accelerator Laboratory, Batavia, Illinois 60510, USA*

F. Cerutti, L.S. Esposito, A. Lechner

*CERN, Geneva 23, CH-1211, Switzerland*

## Abstract

A detailed model of the High Luminosity LHC inner triplet region with new large-aperture $Nb_3Sn$ magnets, field maps, corrector packages, and segmented tungsten inner absorbers was built and implemented into the FLUKA and MARS15 codes. In the optimized configuration, the peak power density averaged over the magnet inner cable width is safely below the quench limit. For the integrated luminosity of 3000 $fb^{-1}$, the peak dose in the innermost magnet insulator ranges from 20 to 35 MGy. Dynamic heat loads to the triplet magnet cold mass are calculated to evaluate the cryogenic capability. In general, FLUKA and MARS results are in a very good agreement.

---

[*] mokhov@fnal.gov

## I. INTRODUCTION

The Large Hadron Collider (LHC) was operated at 4 TeV per beam and 70% of nominal luminosity in 2012. After the consolidation of the accelerator, it will provide 300 fb$^{-1}$ of integrated luminosity at center-of-mass energy of 13-14 TeV by 2021. Subsequently, CERN is planning to make a high luminosity upgrade (HL-LHC) to get 3000 fb$^{-1}$ in 10 years [1], or 4000 fb$^{-1}$ as the ultimate target.

One essential objective of the upgrade is to reduce β* down to 10-15 cm by means of stronger and larger aperture low-beta triplet quadrupoles in the high luminosity Insertion Regions (IRs). The envisaged solution [2] relies on the new Nb$_3$Sn technology, which allows a more compact layout and ~30% higher performance with respect to NbTi coils, and on a 150 mm aperture, doubling the present one of 70 mm. In addition, a superconducting D1 separation dipole will replace the normal-conducting version, as well as new quadrupoles in the Matching Section are foreseen, still based on NbTi technology, but with a larger aperture.

From the quench stability and radiation damage points of view, these magnets should cope with an exceptionally high luminosity. They need to be designed to operate at L = 5×10$^{34}$ cm$^{-2}$ s$^{-1}$ (corresponding to 5 times nominal LHC peak luminosity) or at an ultimate L = 7.5×10$^{34}$ cm$^{-2}$ s$^{-1}$, with appropriate safety margins. Assumed design limits [3-5] are 13 and 4 mW/cm$^3$ for Nb$_3$Sn and NbTi coils, respectively, including a safety factor of 3 on expected quench limits. For long term radiation damage, a tentative dose limit is set to a few tens of MGy, mainly because of the degradation of the epoxy resin used to impregnate Nb$_3$Sn coils [6]. As the first studies of radiation loads in the LHC upgrades have shown [3, 7], one could provide the operational stability and adequate lifetime of the IT superconducting magnets by using tungsten-based inner absorbers in the magnets. Another constraint is given by the total heat power to be evacuated from the ensemble of the Inner Triplet (IT), Corrector Package (CP) and D1 magnets by the cryogenic equipment.

This paper is divided in five parts. The first one is devoted to characterization of the pp-collisions at the LHC interaction points as a source of irradiation of the IR magnets as well as to the approach used to design their protection via inner absorbers. The next part gives the details of the FLUKA and MARS models built to study the problem. Results of Monte Carlo calculations of 3D distributions of power density in the coils of the IT magnets are presented in the third part. The FLUKA and MARS predictions - being in an excellent agreement – are safely below the quench limits. The dynamic heat loads on the magnets are also presented in the section. Results on the quantities related to the radiation damage and lifetime of the IT magnet components – considered as ones of the most critical in the design of the HL-LHC IT magnets - are described in detail in the next part. The fifth part is devoted to the dependencies of the critical radiation-induced quantities on the details of the IT magnet design and their relation to a possible design evolution.

## II. COLLISION DEBRIS AND TRIPLET MAGNETS

### A. Characterization of the radiation source

Proton-proton inelastic collisions taking place in the LHC inside its four detectors generate a large number of secondary particles, on average about 100 (120) per a single proton-proton interaction of 3.5 (7) TeV beams, as calculated with DPMJET-III [8]. There are substantial fluctuations over different events. Moving away from the interaction point (IP), this multiform population evolves, even before touching the surrounding material, because of the decay of unstable particles (in particular neutral pions decaying into photon pairs). Figure 1 illustrates the composition of the debris at 5 mm from the point of a 14 TeV center of mass collision, featuring a ~ 30% increase in number of particles, due to aforementioned decays, and a clear prevalence of photons (almost one half) and charged pions (~ 35%).

Most of these particles are intercepted by the detector and its forward region shielding releasing their energy within the experimental cavern. However, the most energetic ones, emitted at small angles with respect to the beam direction, travel farther in the vacuum and reach the accelerator elements, causing a significant impact on the magnets along the Insertion Regions, in particular the final focus quadrupoles and the separation dipole. Figure 1 shows also the breakdown of the debris components going through the aperture of the TAS (Target Absorber Secondaries) absorber, a protection element consisting of a copper core 1.8-m long located at 20 m from the IP and representing the interface between the detector and the accelerator. The TAS absorbers are installed at each side of the high-luminosity detectors, ATLAS in P1 and CMS in P5. Their protection role, in fact limited to the first quadrupole, is not needed for luminosities up to $0.2 \times 10^{34}$ cm$^{-2}$ s$^{-1}$, which is the upgrade target of LHCb [9].

Despite the fact that the number of particles per collision leaving the TAS aperture is more than one order of magnitude lower than the total number of debris particles, they carry about 80% of the total energy, implying that 40% of the released energy at the IP exits on each side of the experiments. At the nominal HL-LHC luminosity ($5 \times 10^{34}$ cm$^{-2}$ s$^{-1}$), this represents about 3800 W per side that is inevitably impacting the LHC elements and consequently dissipated in the machine and in the nearby equipment (e.g. electronics, racks,...) and in the tunnels walls.

It is fundamental to study how these particles are lost in order to implement the necessary protections for shielding sensitive parts of the LHC magnets and the machine. For these purposes, Monte Carlo simulations of the particle interaction with matter play an essential role, relying on a detailed implementation of physics models and an accurate 3D-description of the region of interest.

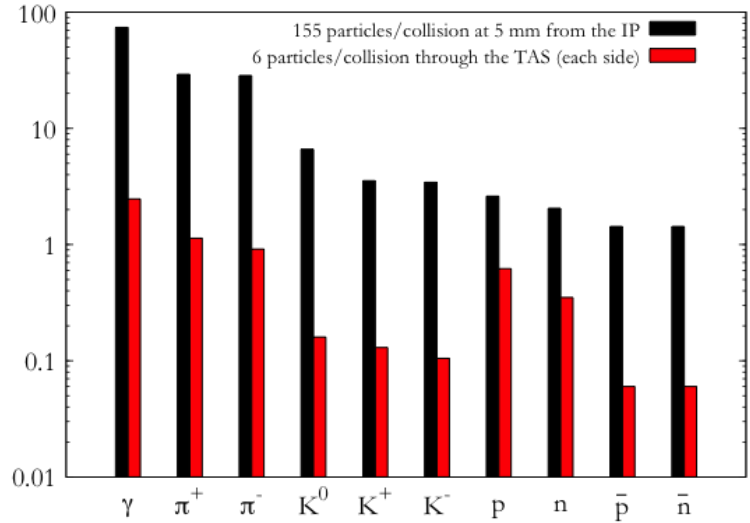

FIG. 1. Number of debris particles per single proton-proton inelastic interaction at 5 mm from the interaction point (black histogram) and at the exit of each 60-mm TAS aperture (red).

### B. Large aperture $Nb_3Sn$ magnets and inner shielding

The LHC upgrade includes replacement of the IP1/IP5 inner triplet 70-mm NbTi quadrupoles with the 150-mm coil aperture $Nb_3Sn$ quadrupoles along with the new 150-mm coil aperture NbTi dipole magnet and orbit correctors. Moreover, a corrector package that includes a skew quadrupole and eight high-order magnets (from sextupole to dodecapole, normal and skew, based on the NbTi technology) will be located between the triplet and the D1.

An octagonal stainless steel beam screen, equipped with 6-mm tungsten absorbers on the mid-planes, is placed inside the cold bore all along the triplet, the CP and the D1, except in Q1 (up to ~32.5m from IP) where the tungsten thickness is increased to 16 mm, compatible with the relaxed aperture requirements. The absorbers are in between the beam screen and the 1.9-K beam pipe: they are supported by the beam screen, and thermally connected to it, whereas they have negligible contact with the cold mass. Therefore, from the point of view of energy deposition, the beam screen function is two-fold:

(i) It shields the coils from the debris by reducing the energy deposited in there,

(ii) It removes a sizable part of the heat load from the 1.9 K cooling system, collecting it at higher temperature.

The present HL-LHC layout foresees six cryostats on each side of the IP: four for the triplet quadrupoles (Q1, Q2A, Q2B and Q3), one for the CP and the last for the D1 dipole. The distance between the magnets in the interconnect regions is 1.5 - 1.7 m (as preliminarily assumed in this study) and an interruption of the beam screen is necessary therein. As a reasonable baseline, we adopt here a 500-mm interruption of the tungsten absorbers in the middle of the interconnect regions.

## III. MONTE CARLO MODELING WITH FLUKA AND MARS CODES

To design such a system in a consistent and confident way, coherent investigations have been undertaken with two independent Monte Carlo codes benchmarked up to the TeV energy region and regularly used in such applications: FLUKA at CERN [10, 11] and MARS15(2014) at Fermilab [12-14]. The studies were done for 7+7 TeV pp-collisions at the luminosity of $5\times10^{34}$ $cm^{-2}s^{-1}$ with a 0.295 mrad half-angle vertical crossing in IP1 (which was found earlier to be the worst case) using DPMJET-III as the event generator.

An identical, very detailed geometry model was created and used in both the codes with same materials and magnetic field distributions in each of the components of the 80-m region from IP through the D1 dipole. Figs. 2 thru 6 show a 3D view of the model and details of the inner parts of the quadrupoles, orbit correctors and dipole D1.

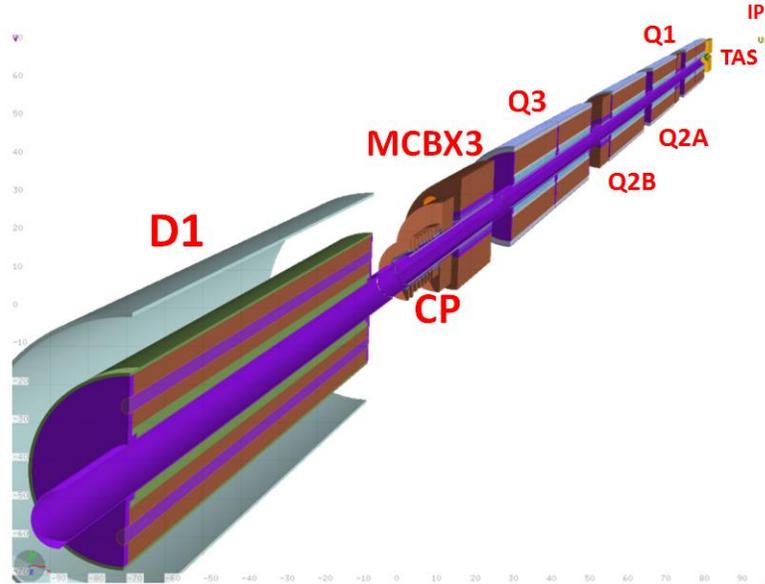

FIG. 2. Computer model of HiLumi LHC inner triplet with correctors MCBX/CP and D1 dipole.

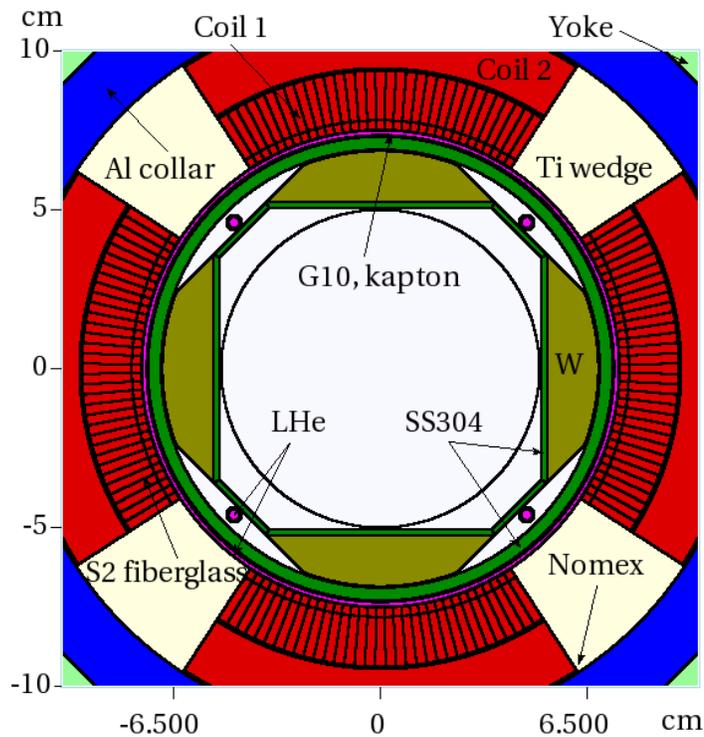

FIG. 3. Details of the FLUKA-MARS model in the innermost region of the Q1 quadrupole. The major difference between Q2-Q3 and Q1 is that the tungsten liner in the former significantly thinner than that in the latter. The Nb$_3$Sn coils are homogenized; inner coils are sub-divided azimuthally and radially for scoring. Nomex and S2 fiberglass insulating inserts are not included in the model; corresponding labels indicate their locations as discussed in the text for Table 2 below.

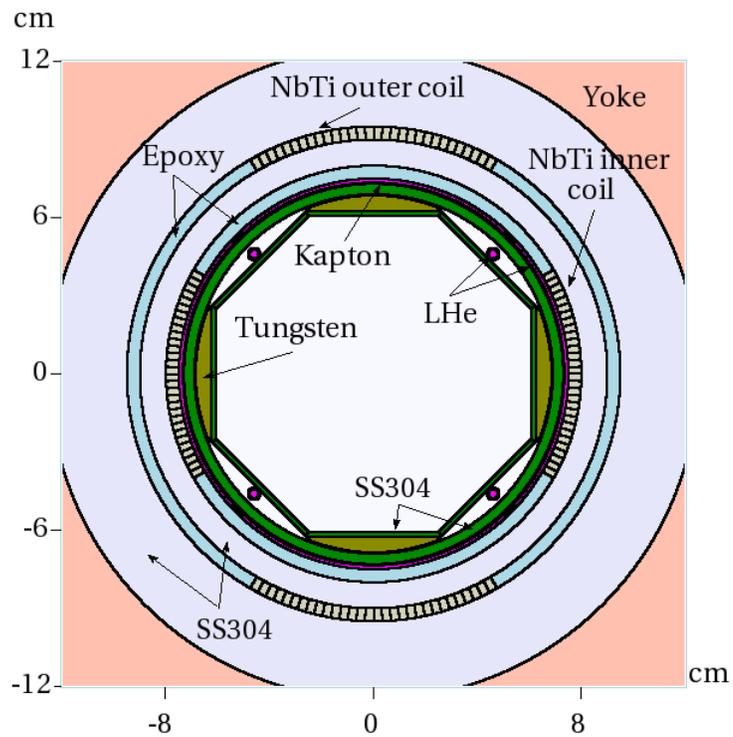

FIG. 4. Cross-sectional view of the FLUKA-MARS model in the central part of the MCBX orbit correctors.

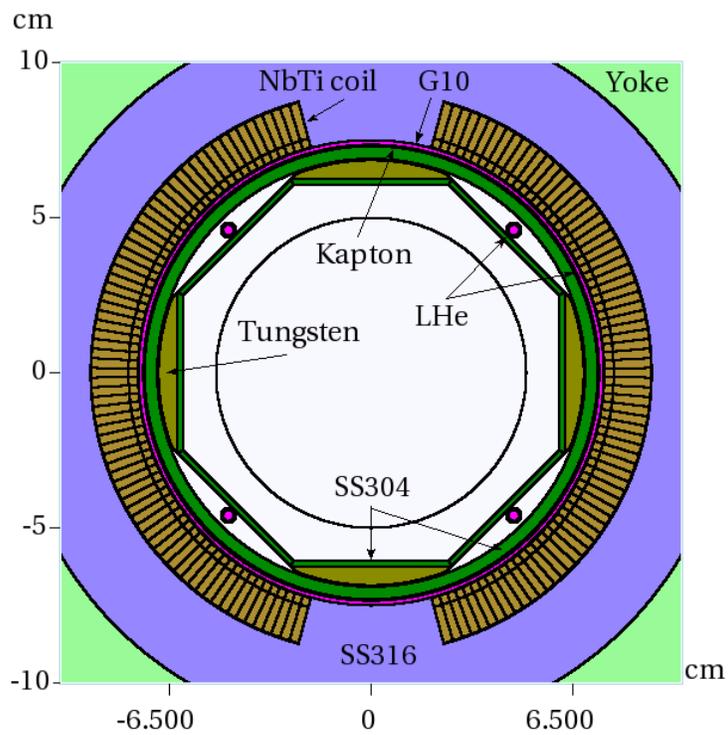

FIG. 5. Cross-sectional view of the FLUKA-MARS model in the central part of the D1 dipole. The coil is homogenized and sub-divided azimuthally and radially for scoring.

Fine-mesh distributions of power density as well as of accumulated for 3000 fb$^{-1}$ integrated luminosity (~10-12 years of HiLumi LHC operation) absorbed dose, neutron fluence and Displacement-Per-Atom (DPA) along with dynamic heat load in every IT component were calculated with FLUKA and MARS in high-statistics runs. The power density and dynamic heat load results are normalized to the luminosity of $5\times10^{34}$ cm$^{-2}$s$^{-1}$, while all other ones to the 3000 fb$^{-1}$ integrated luminosity. Longitudinal scoring bins are 10 cm, and azimuthal ones are 2°. Radially, power density is scored in the superconducting cable width, while dose, fluence and DPA are scored at the azimuthal maxima within the innermost layer equal to 3-mm or its thickness, whatever is thinner.

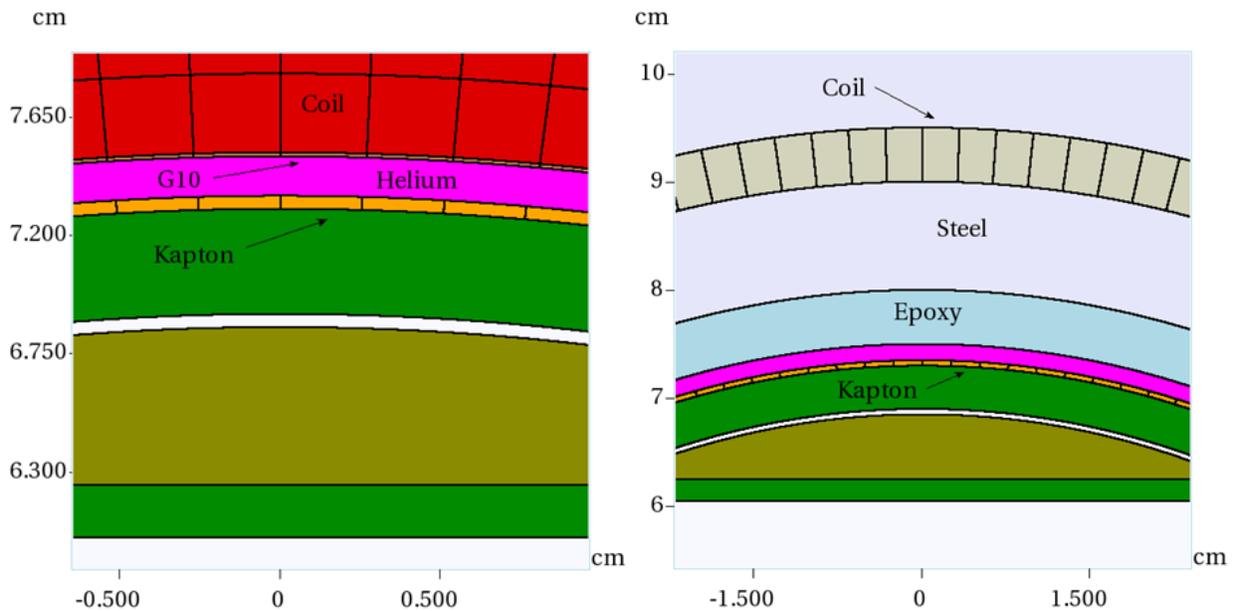

FIG. 6. Cross-sectional view of the FLUKA-MARS model fragments with kapton cells in Q1-Q3 quadrupoles (left) and MCBX correctors (right).

## IV. OPERATIONAL RADIATION LOADS

Power density isocontours at the IP end of the cold mass of the Q2A quadrupole are shown in Fig. 7. The longitudinal peak power density profile on the inner coils of the IT magnets at the azimuthal maxima is presented in Fig. 8. Results from FLUKA and MARS are in an excellent agreement. The peak value in the quadrupoles, 2 mW/cm$^3$, is 20 times less than the assumed quench limit of 40 mW/cm$^3$ in Nb$_3$Sn coils. The peak value of ~1.5 mW/cm$^3$ in the NbTi based coils of the correctors and D1 dipole is almost ten times less of the known quench limit 13 mW/cm$^3$ in such coils.

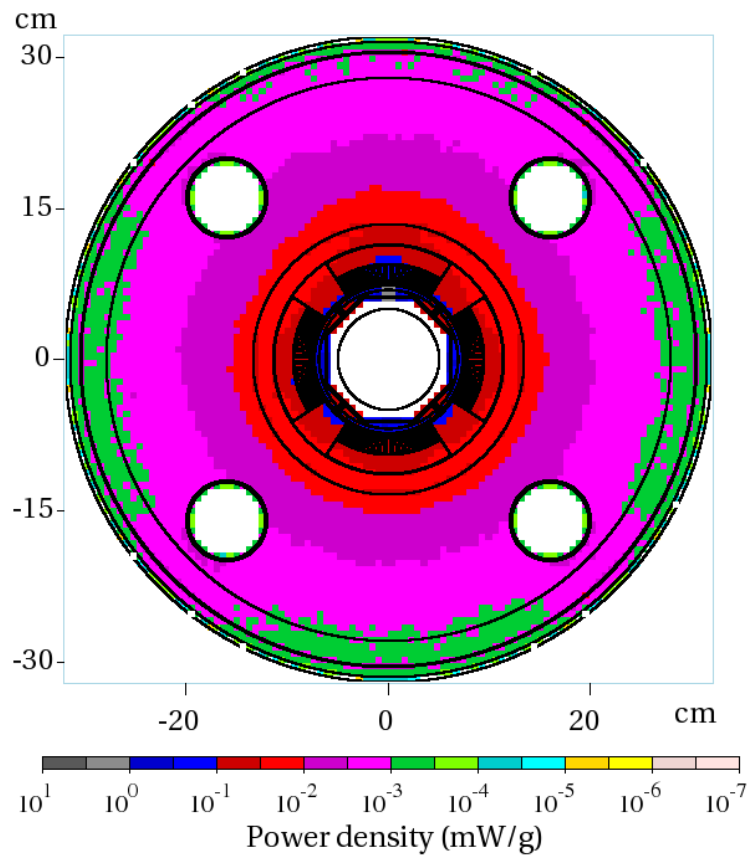

FIG. 7. Power density isocontours at the IP end of Q2A.

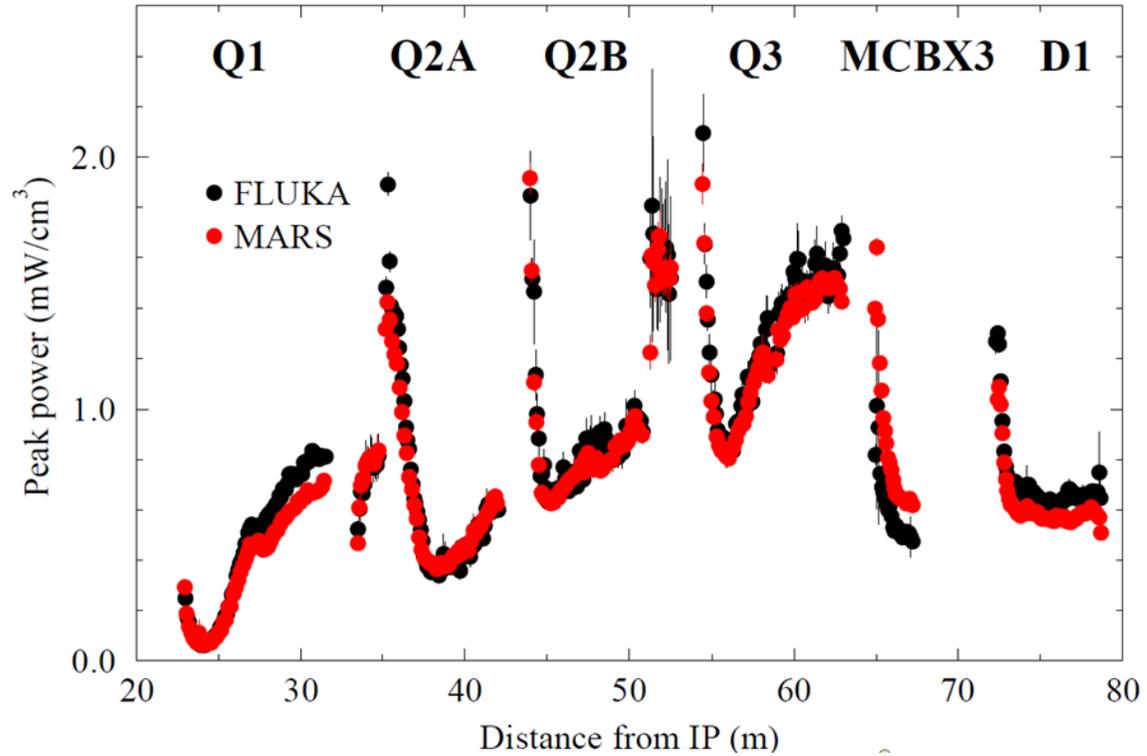

FIG. 8. Longitudinal peak power density profile on the inner coils of the IT magnets.

The total power dissipation in the IT region from the IP1 collision debris splits approximately 50-50 between the cold mass and beam screen with the tungsten absorber: 630 W and 615 W, respectively, from the FLUKA calculations. Total heat load to various components of the inner triplet, including comparison between FLUKA and MARS data, is presented in Table I. One can see that, as far as total heat load is concerned, the two codes agree within about 2%. For the 45-m effective length of the cold mass, the average dynamic heat load on it is ~14 W/m.

TABLE I. Integral power dissipation (W) in components of inner triplet calculated with FLUKA and MARS codes for two different interconnect (IC) gap lengths.

| Component | FLUKA | | | | MARS | |
|---|---|---|---|---|---|---|
| | 10 cm gap in ICs | | 50 cm gap in ICs | | 50 cm gap in ICs | |
| | Magnet cold mass | Beam screen | Magnet cold mass | Beam screen | Magnet cold mass | Beam screen |
| Q1A+Q1B | 100 | 170 | 100 | 170 | 95 | 170 |
| Q2A+orbit corrector | 95 | 60 | 100 | 65 | 100 | 65 |
| Q2B+orbit corrector | 115 | 80 | 120 | 80 | 115 | 80 |
| Q3A+Q3B | 140 | 80 | 140 | 80 | 135 | 75 |
| Corrector package | 55 | 55 | 60 | 55 | 60 | 65 |
| D1 | 90 | 60 | 90 | 60 | 90 | 55 |
| Interconnects | 20 | 140 | 20 | 105 | 15 | 85 |
| Total | 615 | 645 | 630 | 615 | 615 | 600 |

## V. LIFETIME RADIATION LOADS

The peak dose and DPA – the quantities that define radiation damage and lifetime of insulators and non-organic materials of the IT magnets, respectively – are calculated at the azimuthal maxima in the innermost tiny layers of each the IT component shown in Figs. 3-6.

The longitudinal peak dose profiles on the inner coils and insulating materials are presented in Fig. 9. The values in the MCBX orbit correctors in the Q1-Q2A, Q2B-Q3 and Q3-D1 regions are given for the epoxy layer (FLUKA) and kapton layer (MARS); see Figs. 4-6 for details. Results from FLUKA and MARS are again in a good agreement. The larger aperture IT magnets and the tungsten absorbers implemented perform very well, reducing the peak values of both power density and absorbed dose in the HiLumi LHC IT to the levels which correspond to the LHC nominal luminosity.

The integrated peak dose in the IT magnet insulation reaches 30-36 MGy in the MCBX3 corrector, 28-30 MGy in the quadrupoles and ~22 MGy in the D1 dipole. This is at the common limits for kapton (25-35 MGy) and CTD-101K epoxy (25 MGy) or slightly above them. The maximum peak dose in the coils is about 25 MGy for quadrupoles and ~15 MGy for the D1 dipole.

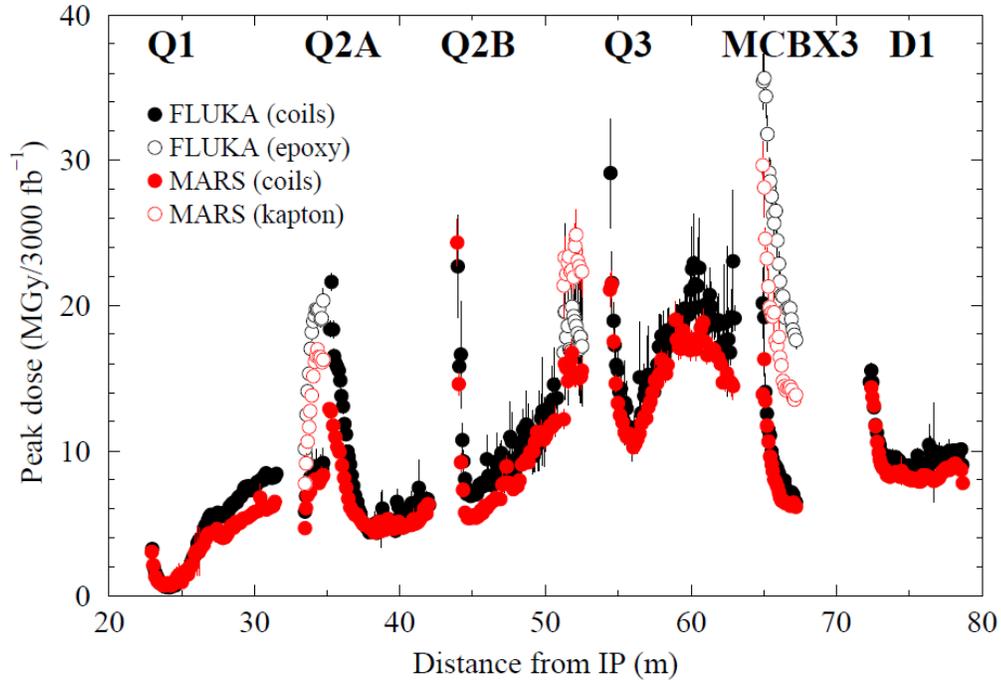

FIG. 9. Longitudinal peak dose profile on inner coils and nearby insulators.

Another important issue is related to absorbed dose at the level of the beam screen. In order to mitigate electron cloud heating, a thin-film carbon coating is deposited on the inner surface. Mechanical stability of the coating depends on various factors including irradiation. The calculated peak absorbed dose in the stainless steel beam screen is shown in Fig. 10.

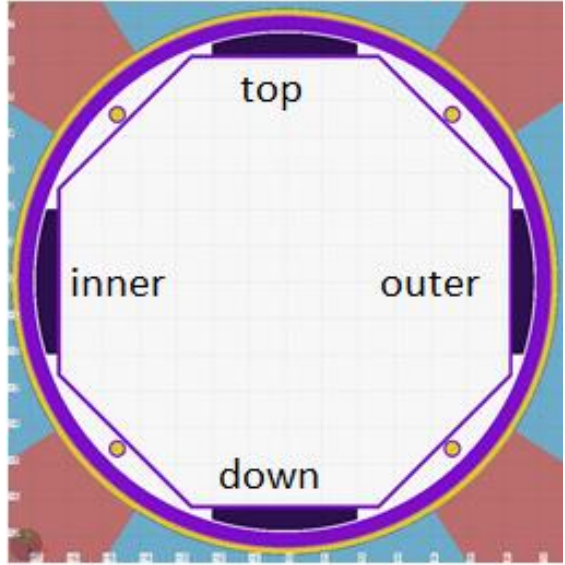

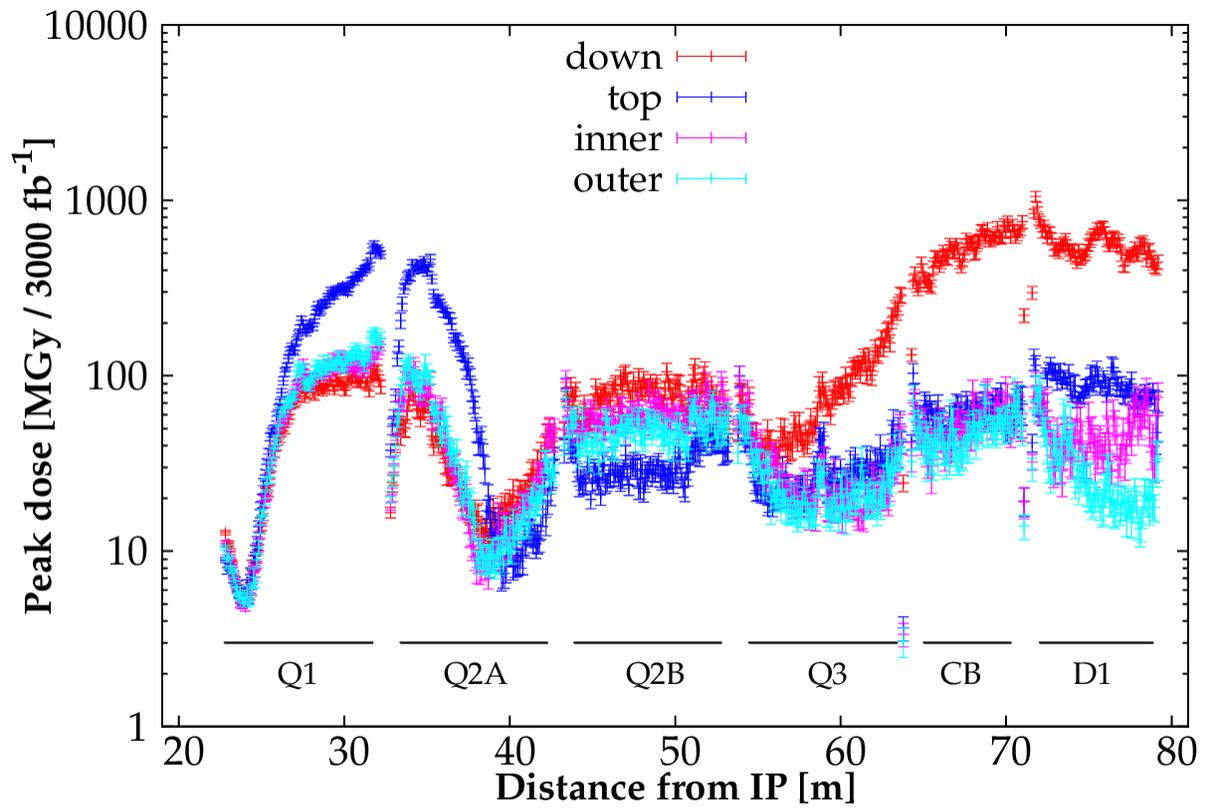

FIG. 10. Longitudinal peak dose profile (bottom) on different segments of the beam screen (top).

Table II summarizes the peak predicted absorbed dose in the hottest components of the inner triplet. One can see that in the hottest spots of the triplet the calculated absorbed dose for the target integrated luminosity is near or slightly above the lifetime limit which means that degradation of material properties becomes relevant. One has to point out that the real magnets are more complicated than the simulation models built for this study. For example, some insulating inserts made of S2 fiberglass and Nomex as shown, e.g., in Fig. 3, were not included in the model. Nevertheless, to estimate the radiation loads to these components slightly beyond the current model, Table II includes the extrapolated peak dose values in such inserts using the calculated spatial dose gradients. That is why some peak values in Table II do not correspond to the data shown in Fig. 9.

Table II. Predicted radiation load (MGy) to organic materials in hottest components of the inner triplet.

| Component | Common name | Material | Maximum calculated value per $I_0=3000$ fb$^{-1}$ | Limit |
|---|---|---|---|---|
| Q2B | Insulation | Kapton | 34 | 25-35 |
| Q2B | Insulation | G10 | 25 | 20 |
| Q2B | Insulation/Glue | Epoxy CTD-101K | 24 | 25 |
| Q2B | Insulation | S2 fiberglass | 24 | 15 |
| Q2B | Insulation | G11 | 24 | 25-40 |
| Q2B | Support material | Nomex | 6.7 | 15 |
| Q2B | Insulation | Polyimide | 6.7 | 25 |
| MCBX3 | Insulation | Kapton | 30 | 25-35 |
| MCBX3 | Insulation/Glue | Epoxy CTD-101K | 27 | 25 |
| D1 | Insulation | Kapton | 22 | 25-35 |
| D1 | Insulation | G10 | 20 | 20 |

Degradation of the critical properties of inorganic materials of the IT magnets – $Nb_3Sn$ and NbTi superconductors, copper stabilizer and mechanical structures – is usually characterized not by absorbed dose but by integrated neutron fluence and by DPA accumulated in the hottest spots over the expected magnet lifetime. DPA is the most universal way to characterize the impact of irradiation on inorganic materials. In both FLUKA and MARS, all products of elastic and inelastic nuclear interactions as well as Coulomb elastic scattering (NIEL) of transported charged particles (hadrons, electrons, muons and heavy ions) from ~1 keV to TeV energies contribute to DPA using energy-dependent displacement efficiency. For neutrons at <20 MeV (FLUKA) and <150 MeV (MARS), the ENDF-VII database with NJOY99 processing is used in both the codes.

The longitudinal peak neutron fluence and peak DPA profiles on the IT magnet coils are presented in Fig. 11. The peaks are generally observed at the inner coils; therefore, the data is given there. With the vertical crossing in IP1, the MCBX3 orbit corrector is the exception with the peak in the outer coil in the vertical plane (see Fig. 4). To see this effect, the MARS data in Fig. 11 for MCBX3 is given for the outer coil, while FLUKA shows results for the inner coil as in all other magnets.

Contrary to the power density and dose distributions driven by electromagnetic showers initiated by photons from neutral pion decay, DPA peaks at the non-IP end of the Q1B quadrupole. At that location, about 70% of DPA is from neutrons with kinetic energy below 20 MeV, ~25% from transported nuclear recoils with the energy above 0.25 keV per nucleon, and the rest is due to other transported particles and non-transported recoils. One can also see from Fig. 11 that a quite definite scaling is observed between the values of peak neutron fluence and peak DPA.

FLUKA and MARS results on neutron fluence are in a quite good agreement. Results on DPA from two codes are also very similar in the Q1A through Q3B region with the MARS's ones being slightly higher than those from FLUKA. At the same time discrepancy in DPA prediction increases in the opposite direction in D1, despite the consistency of neutron fluence values. Our attempts to explain this effect in the DPA behavior at the very end of the studied region have been unsuccessful so far.

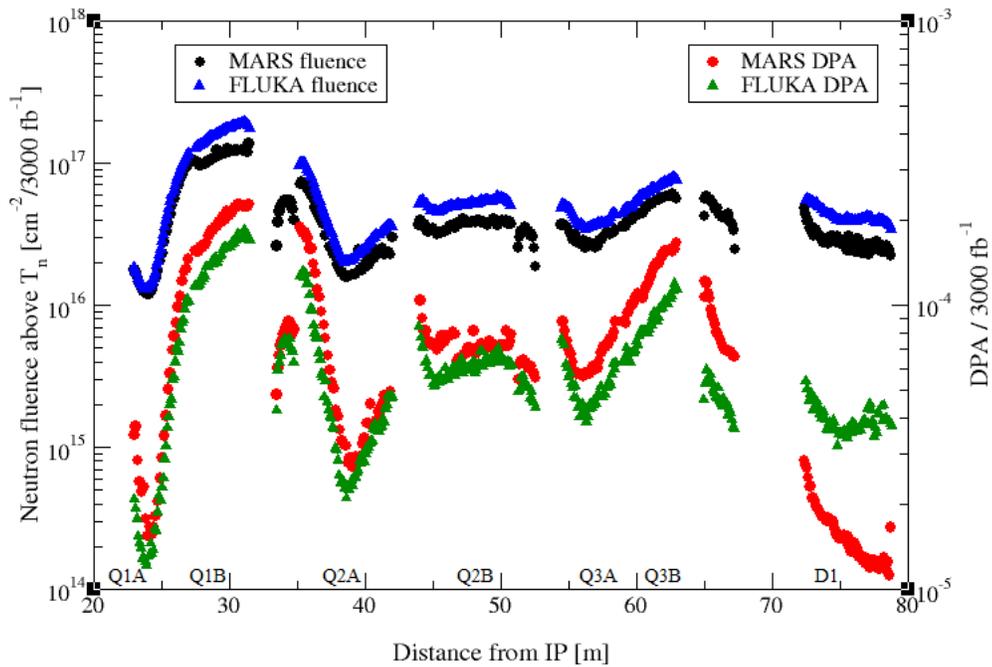

FIG. 11. Longitudinal peak neutron fluence and peak DPA profiles along the hottest regions in the IT magnet coils.

The peak in the Q1B inner coil is about $2\times10^{-4}$ DPA per 3000 $fb^{-1}$ integrated luminosity. In other IT components it is about $(1\pm0.5)\times10^{-4}$. These numbers should be acceptable for the superconductors and copper stabilizer provided periodic annealing during the collider shutdowns. Taking into account a good correlation of DPA with neutron fluence in the coils, one can also compare the latter with the known limits. In the quadrupole coils, the peak fluence is $\sim2\times10^{17}$ $cm^{-2}$ which is substantially lower than the $3\times10^{18}$ $cm^{-2}$ limit used for the $Nb_3Sn$ superconductor. In the orbit corrector and D1 dipole coils, the peak fluence is ~$5\times10^{16}$ cm$^{-2}$ which is again lower than the $10^{18}$ cm$^{-2}$ limit used for the NbTi superconductor.

The integrated DPA in the magnet mechanical structures are 0.003 to 0.01 in the steel beam screen and tungsten absorber, ~ $10^{-4}$ in the collar and yoke, and noticeably less outside. These are to be compared to a ~10 DPA limit for mechanical properties of these materials. Neutron fluences in the IT mechanical structures range from $3\times10^{16}$ cm$^{-2}$ to $3\times10^{17}$ cm$^{-2}$ compared to the $10^{21}$ cm$^{-2}$ to $7\times10^{22}$ cm$^{-2}$ limits.

## VI. CRITICAL DEPENDENCIES AND DESIGN EVOLUTION

The beam screen equipped with tungsten absorbers represents the backbone element for the protection of the IT magnets. Therefore, the details of its design play a crucial role in determining its actual effectiveness.

After the preliminary studies described in the previous section, new estimates were necessary to include:

− the real absorber material, INERMET 180 that has a density of 18 g cm$^{-3}$, about 8% less than pure tungsten, implying a reduced shielding performance,

− the first prototype drawing [15] that takes into account the machinability of INERMET and the required size of the cooling tubes as dictated by preliminary cryogenics estimates,

− the reduction of the beam screen thickness (from 2 to 1 mm) necessary to let the structure respond elastically to possible deformations occurring during a quench.

Figure 12 (left) shows the transverse section of the beam screen model (BS#2) embedding the aforementioned modifications. It can be compared to the model (BS#1) used in the calculations reported in the previous section (see Figs. 3-6). The longitudinal peak dose profile on the inner coils of the IT magnets is presented in Fig. 13 (left) for the case of BS#1 (black points) and of

BS#2 (red points). In the latter case, the accumulated peak dose turns out to be systematically higher all along the IT magnets, almost doubling its value in the Q3 and reaching about 55 MGy in the MCBX3 corrector (however in an azimuthal position that in the MCBX design being detailed in the meantime lies in the collar outside the coil region). Along the Q2, most of the impacting debris, positively charged, is pushed by the magnetic field from the crossing angle side to the opposite one, i.e. from top to bottom in the assumed crossing scheme, where the outgoing beam points upwards. This moves the energy deposition peak through different azimuthal regions, which in the revised design (BS#2) are no longer shielded by the beam screen absorbers, hence yielding the resulting substantial increase. In order to address this drawback, we considered a third version of the beam screen (BS#3), where the INERMET absorbers were extended as much as possible to cover the coils towards the poles (see Fig. 12, right). The estimated peak dose distribution (green points in Fig. 13, left) shows a significant improvement in the Q3-CP region, when compared to the BS#2 case. It should also be noted that, mainly due to the reduced absorber density, the sharing of the total deposited power between the cold mass and beam screen gets unbalanced, moving to 55-45 and making the heat released in the cold mass approach 700W (at $5 \times 10^{34}$ cm$^{-2}$ s$^{-1}$).

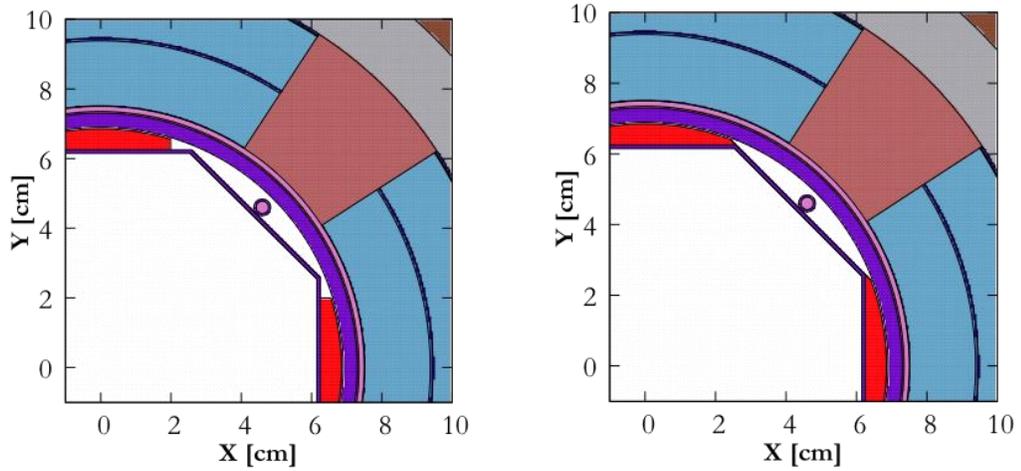

FIG. 12. Left: beam screen model as per the first conceptual design (BS#2) [15]. Right: beam screen model with the modification of the absorbers driven by energy deposition considerations (BS#3).

Another crucial aspect is the longitudinal interruption of the beam screen and its absorbers, which is necessary between two consecutive cryostats in order to host a bellow and a BPM. As mentioned in the previous section, we initially assumed a 500-mm gap. Shorter gaps are possible if the BPMs are going to be equipped with absorber layers like the ones in the beam screen. To mimic this case, we looked at the effect of a 100-mm gap, which should be considered as the most optimistic case. The peak dose dependence on the gap length is presented in Fig. 13 (right) where the improvement achieved downstream the Q2A-Q2B, Q2B-Q3 and especially Q3-CP interconnects is visible, with a reduction from 55 to 35 MGy in the MCBX3 for the BS#2 design. Therefore, the actual implementation of the absorber layers in the design of both the beam screen and the relevant BPMs considerably affects the maximum dose expected in the IT coils.

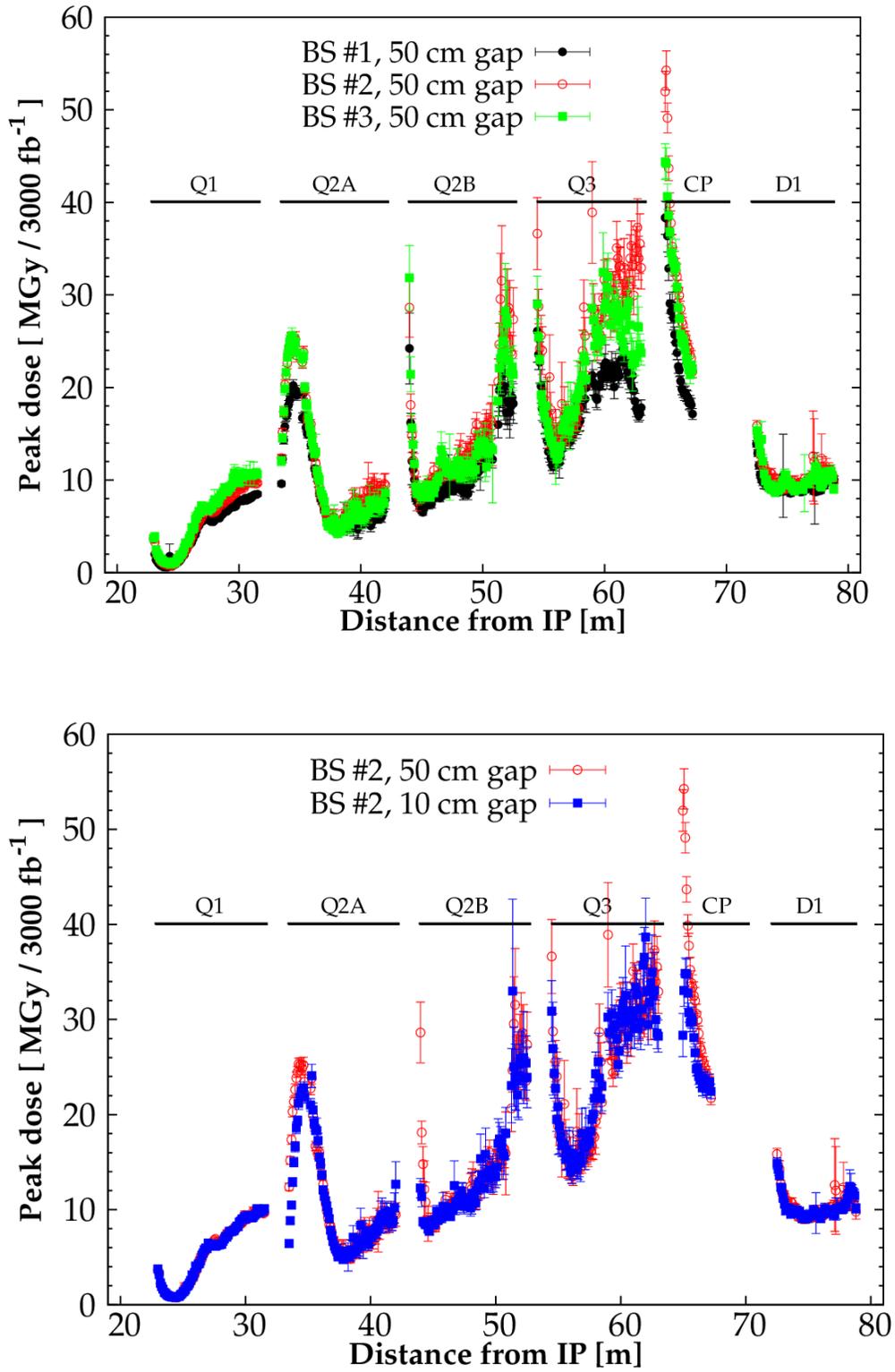

FIG. 13. Longitudinal distributions of peak dose on the inner coils of the IT magnets referring to different beam screen designs (top) and to different lengths of the beam screen gap in the interconnects (bottom).

## VII. CONCLUSIONS

It was shown that 80% of the energy released in pp-collisions leak through the TAS apertures on the two sides of the experiments, resulting in 3.8-kW dynamic heat load impacting the HL-LHC accelerator components on each side around IP1 and IP5. Very detailed descriptions of geometry, materials and magnetic fields for all the components in the inner triplet regions were implemented into the FLUKA and MARS15 models to find the optimal parameters of the protective components needed to assure the operational performance of the IP1/IP5 150-mm aperture $Nb_3Sn$ final focus quadrupoles along with the 150-mm aperture NbTi D1 separation dipoles and corrector magnets. Results of simulations with the two independent codes were found to be in a good agreement. It was demonstrated that the proposed system of the tungsten-based inner absorbers assures the quench stability of the IT magnets with a safety margin close to or exceeding a factor of 10 as well as manageable dynamic heat loads on the IT cryogenic system. The peak DPA in non-organic materials of the IT magnet coils is about $2 \times 10^{-4}$ DPA per 3000 $fb^{-1}$ integrated luminosity that should be acceptable for the superconductors and copper stabilizer provided annealing during the collider shutdowns. The peak dose accumulated in the magnet non-organic materials at the same integrated luminosity is close to the established limits. To provide a reasonable safety margin, it implies that the limits need better understanding, advanced radiation-resistant materials should be considered and the ways to further improve protection efficiency of the inner absorber system need a further attention (e.g., a 100-mm gap in the beam screen and its absorbers rather than a 500-mm one). Finally, whereas the study demonstrated the effectiveness of the conceptual design solution corroborated by a robust cross-validation of the calculation tools, detailed numbers should not be taken as references carved in stone, since the actual design is still evolving, with the impact discussed in the previous section.

**ACKNOWLEDGEMENTS**

This work was supported by Fermi Research Alliance, LLC, under contract No. DE-AC02-07CH11359 with the U.S. Department of Energy through the US LARP Program, and by the High-Luminosity LHC Project.